 \documentclass[final,5p,times,twocolumn]{elsarticle}

\usepackage{graphicx}
\usepackage{amsmath}
\usepackage{amssymb}
\newcommand{\mev}{\!\mathrm{MeV}}

\journal{Physics Letters B}

\begin{document}

\begin{frontmatter}



\title{Observation of the Charged Hadron $Z_c^{\pm}(3900)$ and Evidence for the Neutral $Z_c^0(3900)$  in $e^+e^-\to \pi\pi J/\psi$ at $\sqrt{s}=4170$~MeV}


\author[nu]{T.~Xiao}
\author[nu]{S.~Dobbs}
\author[nu]{A.~Tomaradze}
\author[nu]{Kamal~K.~Seth\corref{cor1}}
\ead{kseth@northwestern.edu}

\address[nu]{Northwestern University, Evanston, Illinois 60208, USA}

\begin{abstract}
Using 586~pb$^{-1}$ of $e^+e^-$ annihilation data taken with the CLEO-c detector at $\sqrt{s}=4170$~MeV, the peak of the charmonium resonance $\psi(4160)$, we analyze the decay $\psi(4160)\to \pi^+\pi^- J/\psi$, and report observation of the charged state $Z_c^\pm(3900)$ decaying into $\pi^\pm J/\psi$ at a significance level of $>5 \sigma$.  We obtain $M(Z_c^\pm)=3886\pm4(\text{stat})\pm 2(\text{syst})$~MeV and $\Gamma(Z_c^\pm)=37\pm4(\text{stat})\pm 8(\text{syst})$~MeV, which are in good agreement with the results for this resonance obtained by the BES~III and Belle Collaborations in the decay of the resonance Y(4260).  We also present first evidence for the production of the neutral member of this isospin triplet, $Z_c^0(3900)$ decaying into $\pi^0J/\psi$ at a $3.5\sigma$ significance level.
\end{abstract}

%\begin{keyword}


%\end{keyword}

\end{frontmatter}

The BES III~\cite{besz} and Belle~\cite{bellez} Collaborations have recently reported observation of a charged hadron $Z^{\pm}_c(3900)$ with mass $\sim3900$~MeV which decays into a charged pion and $J/\psi$. This is an important finding because a charged hadron decaying into a charmonium state plus a charged meson must contain at least four quarks. If confirmed, this observation would herald the possible existence of a family of other charged states in this mass region.  
Several charged hadrons were reported earlier by Belle in the charmonium and the bottomonium regions~\cite{belle}, but not all have been independently confirmed, and some remain controversial.

The BES~III observation~\cite{besz} of $Z^{\pm}_c(3900)$ was made in $e^+e^-$ annihilation at the vector resonance Y(4260), which is known to have unusual characteristics, and does not fit in the conventional spectrum of charmonium states. 
The Belle observation~\cite{bellez} was also made at Y(4260), which was populated by initial state radiation from 967~fb$^{-1}$ of $e^+e^-$ annihilations at the $\Upsilon(nS)$ resonances.
Clearly, it is very important to make an independent confirmation of the existence of $Z^{\pm}_c(3900)$, and to also determine if it is populated in $e^+e^-$ annihilation at a resonance which, unlike Y(4260), has well established charmonium credentials, such as $\psi(4160)$, the $2^3D_1$ state of charmonium. 

In this letter we report the results of our search for $Z_c(3900)$~\cite{nuz}. We confirm the BES~III and Belle observation of the charged $Z^{\pm}_c(3900)$, and report evidence for $Z^{0}_c(3900)$, the neutral member of this isospin triplet.

We use 586 $\mathrm{pb}^{-1}$ of $e^+e^-$ collision data taken at $\sqrt{s}=4170~\mev$ at the CESR collider at Cornell University, with final state particles detected and identified in the CLEO-c detector, which has been described in detail elsewhere~\cite{cleodetector}. \\[20pt]

%\noindent
%\textit{Observation of the charged $Z^{\pm}_c(3900)$}

We first present the results for our observation of $Z^{\pm}_c(3900)$ in the same decay chain as BES~III and Belle,
$$e^+e^- \to\pi^\mp Z_c^\pm, ~~ Z_c^\pm\to \pi^\pm J/\psi,$$
but using CLEO-c data taken at $\sqrt{s}=4170$~MeV, on the peak of the well-known $\psi(4160)$, the $2^3D_1$ charmonium resonance.

For the reaction $\psi(4160) \to \pi^+\pi^- J/\psi$, $J/\psi \to \mu^+\mu^-,~e^+e^-$, we select events with 4 charged particle tracks with zero net charge. Tracks are reconstructed in the region with $|\cos\theta_{tr}|<0.93$, where $\theta_{tr}$ is the polar angle, and are required to be well-measured and consistent with originating at the interaction point.

Charged particle tracks are first identified on the basis of their momenta. Leptons (e, $\mu$) from the decay of $J/\psi$ have momenta $>1$~GeV, and pions have momenta $<1$~GeV, which makes $\pi/\text{lepton}$ separation easy. Pion candidates are additionally required to have energy loss in the drift chamber ($dE/dx$)  consistent within $3\sigma$ with that expected for pions.  

Muons are distinguished from the electrons based on the variable $E_{CC}/p$, where $p$ is the track momentum measured in the drift chamber and $E_{CC}$ is the energy deposited in the calorimeter associated with the charged particle track.  This variable cleanly separates electrons, with $E_{CC}/p\approx1$, from muons which have $E_{CC}/p<0.25$.
To reject backgrounds from converted photons, $\gamma\to e^+e^-$, where the resulting electrons are misidentified as pions, we require $\cos(\pi^+,\pi^-)<0.98$ and $\cos(\pi^\pm,e^\mp)<0.98$.

To select fully reconstructed events and improve mass resolution, a 4C kinematic fit is performed constraining the $\pi^+\pi^-J/\psi$ final state to a common vertex with $\chi^2_{\mathrm{vertex}} < 20$, and the $e^+e^-$ collision energy and momentum with $\chi^2_{\mathrm{4C~fit}} < 20$. In the following, we use the momenta of the charged particles after the kinematic fit. To select events containing $J/\psi \to \mu^+\mu^-,~e^+e^-$ decays, we select events with the dilepton mass $M(\mu^+\mu^-)$ and $M(e^+e^-)$ consistent with $M(J/\psi)$ within $\pm12$~MeV.

\begin{figure}[!tb]

\begin{center}
\includegraphics[width=2.4in]{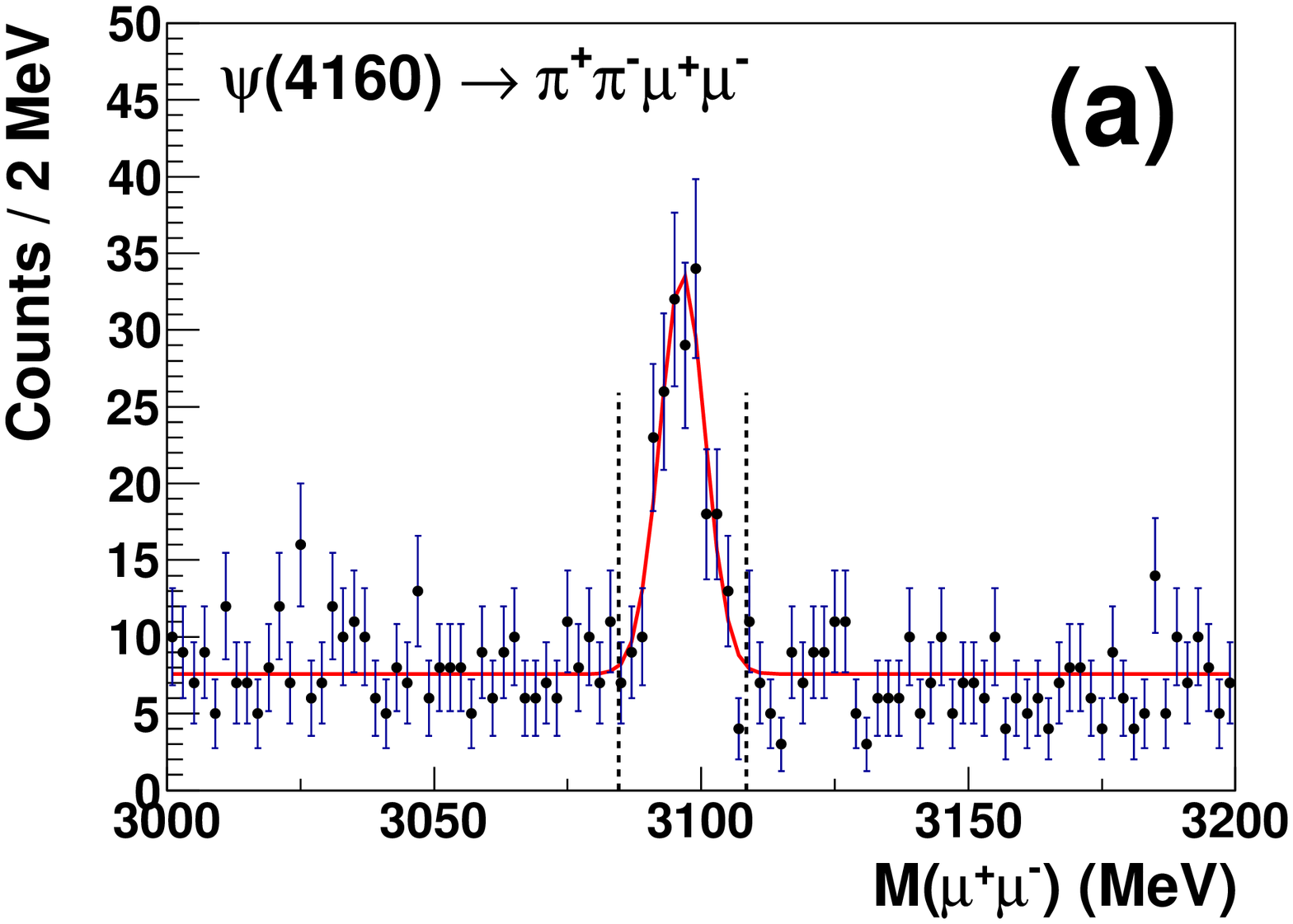}

\includegraphics[width=2.4in]{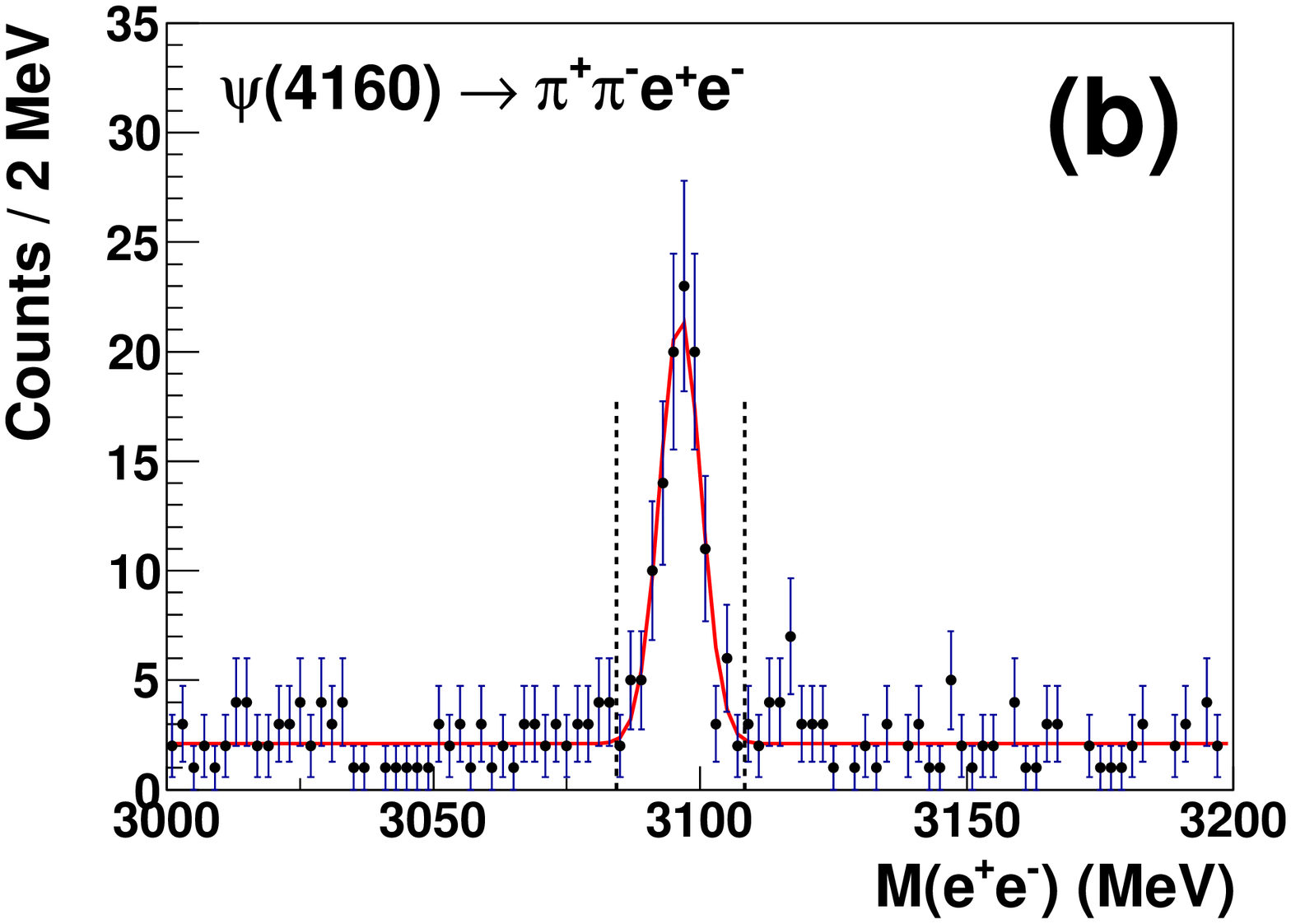}
%\includegraphics[width=2.4in]{chg-ee_jpsi.eps}
%\includegraphics[width=2.4in]{chg/jpsi/ee_jpsi.eps}

%\vspace*{10pt}

%\vspace*{10pt}

\includegraphics[width=2.4in]{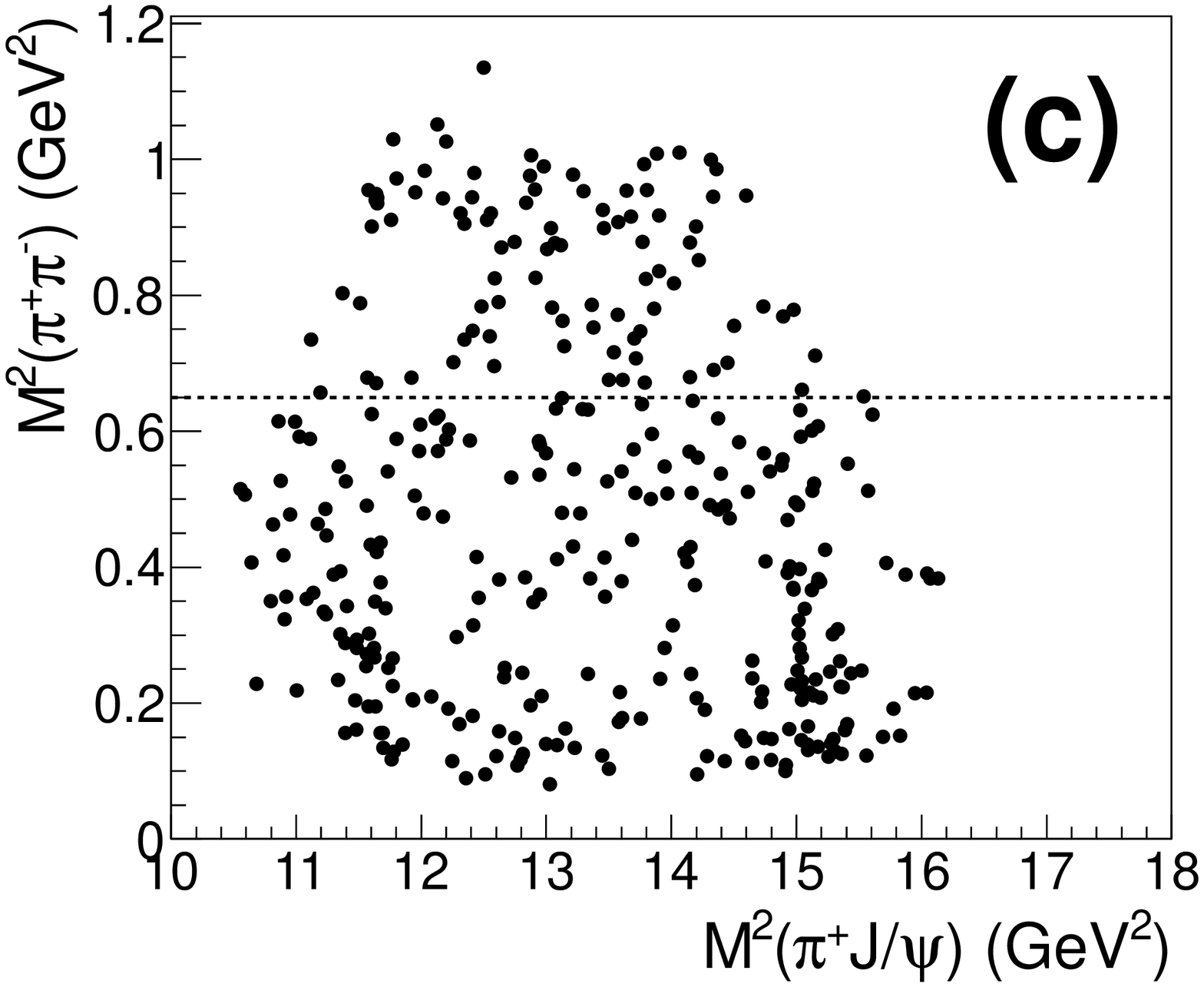}
%\includegraphics[width=2.4in]{chg-2Dleft.eps}
%\includegraphics[width=2.4in]{chg/2Dleft/2Dleft.eps}

%\vspace*{10pt}

\includegraphics[width=2.4in]{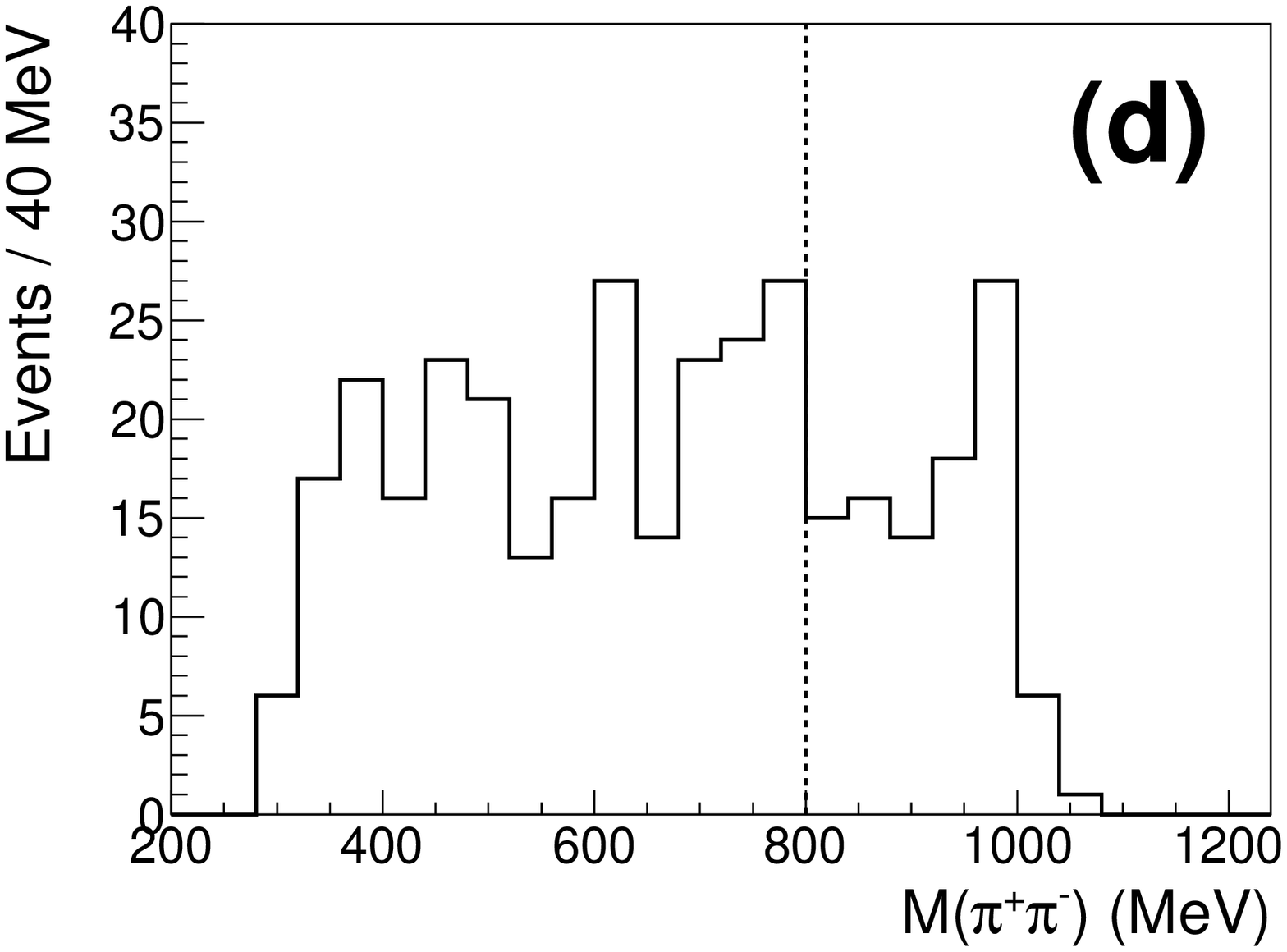}
\end{center}

\caption{(a,b) Invariant mass distributions for $\psi(4160)\rightarrow\pi^+\pi^-(\mu^+\mu^-,e^+e^-)$. The curves are fits as described in the text. The dashed vertical lines define the $J/\psi$ resonance region used in subsequent analysis; (c) Dalitz plot for $M^2(\pi^+\pi^-)$ versus $M^2(\pi^+J/\psi)$; (d) $M(\pi^+\pi^-)$ projection of the Dalitz plot.  The dashed lines in (c) and (d) indicate division of the data into two parts as described in the text.}

\end{figure}

\begin{figure}[!t]

\begin{center}
\includegraphics[width=2.8in]{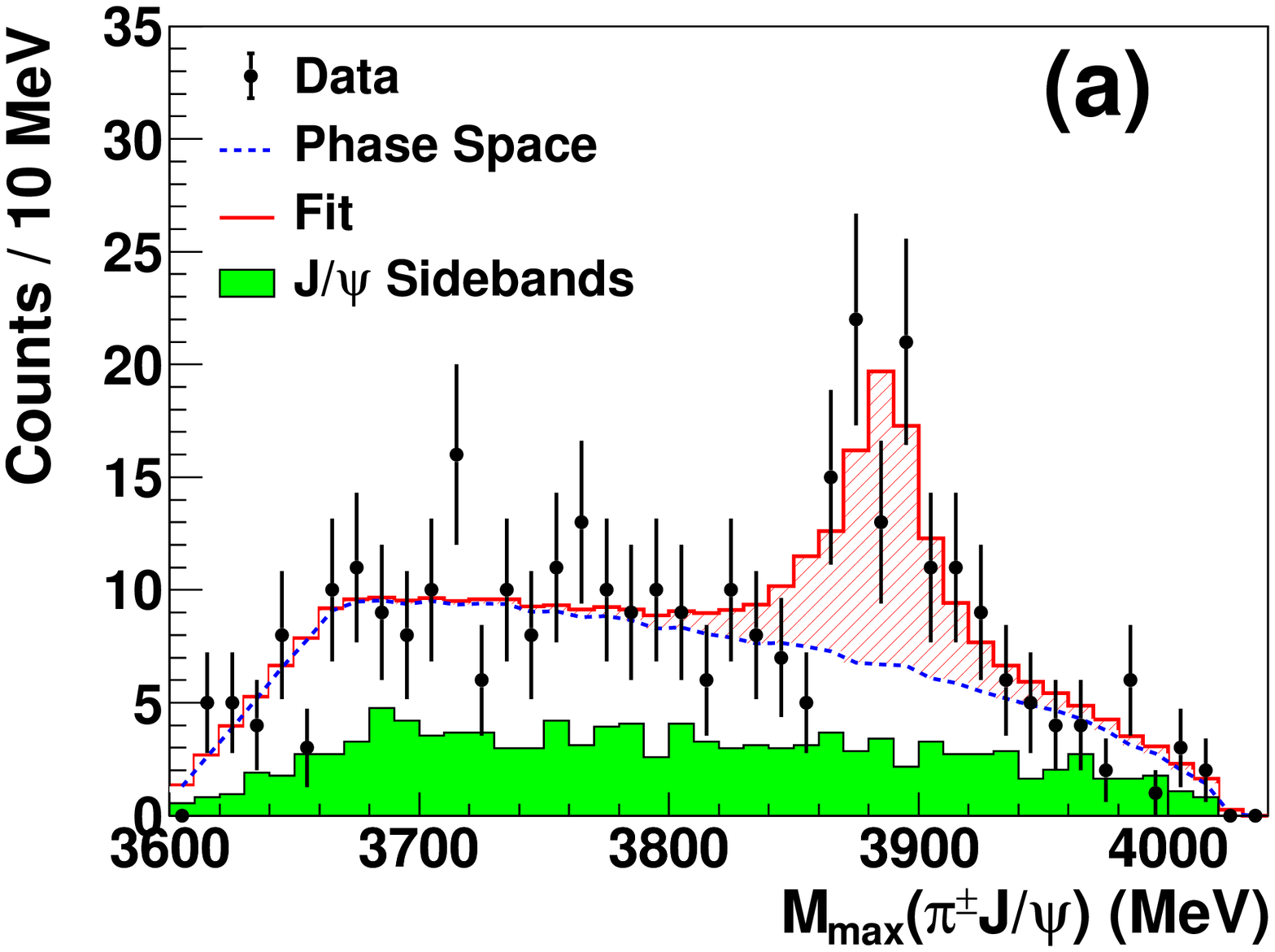}

\includegraphics[width=2.8in]{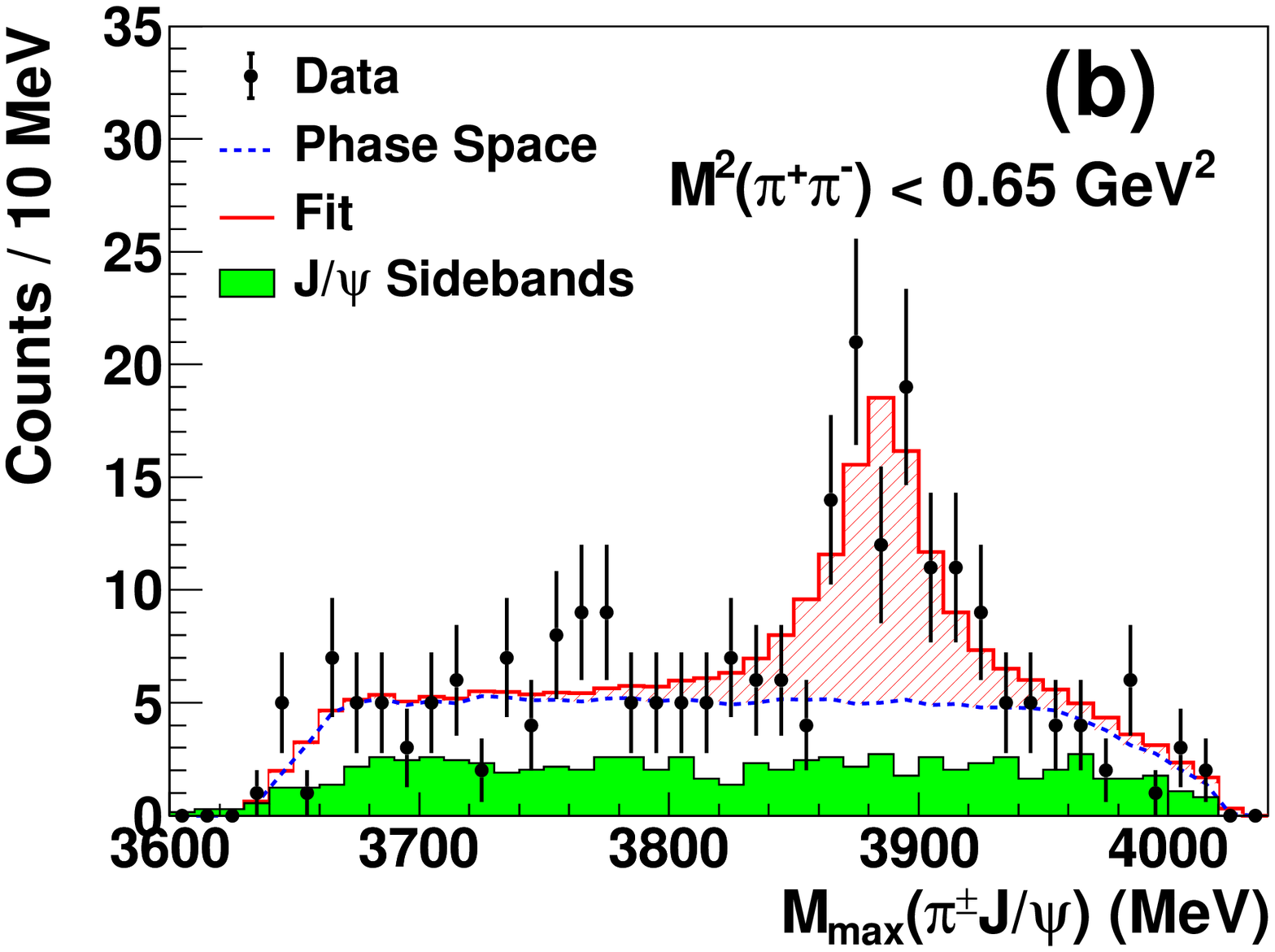}
%\includegraphics[width=2.8in]{chg-max_mass.eps}
%\includegraphics[width=2.8in]{chg/pipi08_max_m/max_mass.eps}
%\includegraphics[width=2.8in]{chg/pipi08_max_m-new/max2_mass.eps}

%\vspace*{10pt}

\includegraphics[width=2.8in]{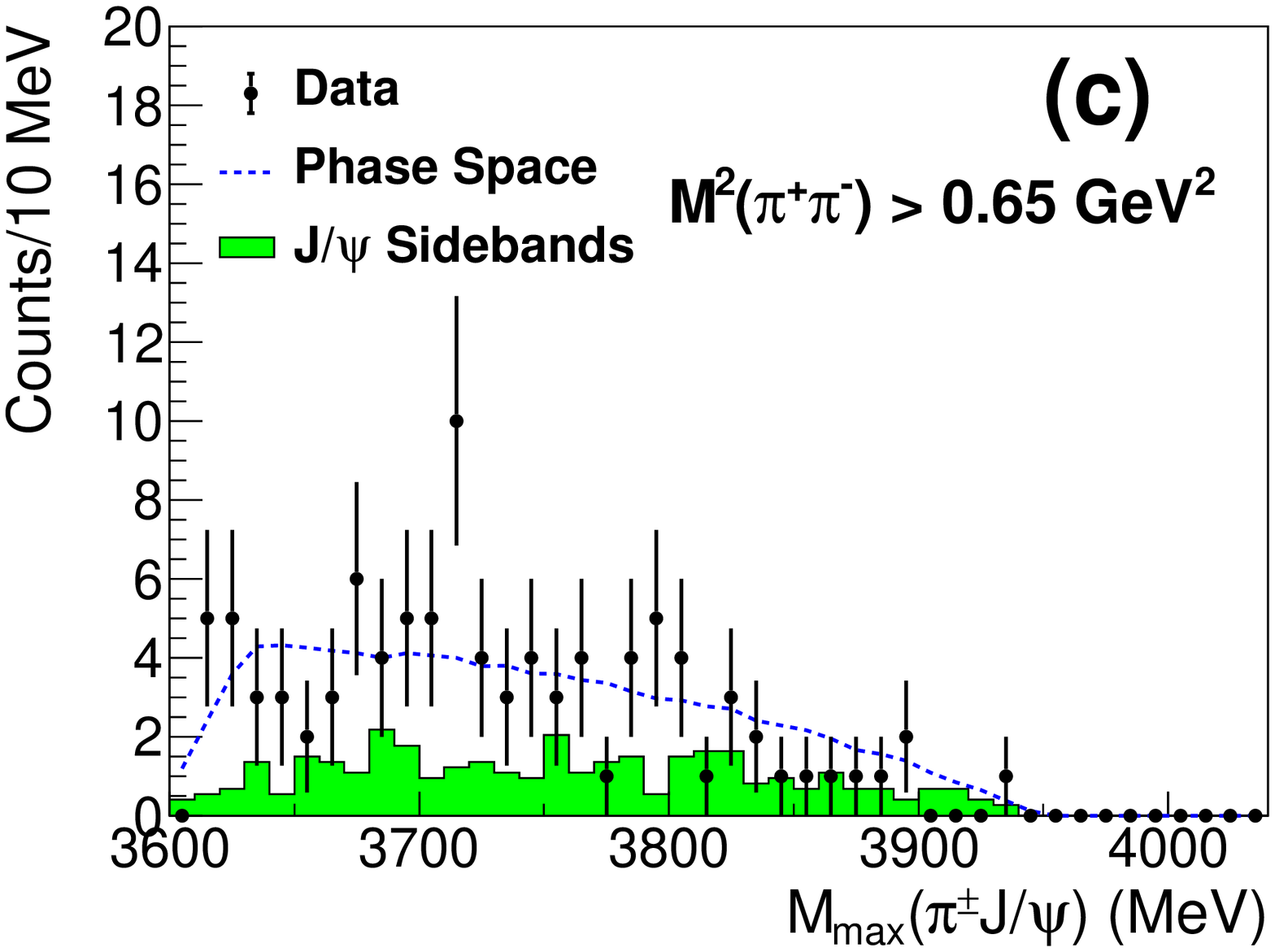}
\end{center}

\caption{Distributions of $M_\text{max}(\pi^{\pm} J/\psi)$ as observed in the decay $\psi(4160)\rightarrow\pi^+\pi^-J/\psi$. The histograms show the fits.  The dashed curves show the MC-determined phase-space background.  The hatched peak shows the contribution of the Breit-Wigner resonance. (a) with no cut in $M^2(\pi^+\pi^-)$, (b) $M^2(\pi^+\pi^-)<0.65$~GeV$^2$, (c) $M^2(\pi^+\pi^-)>0.65$~GeV$^2$.}

\end{figure}

The $J/\psi$ decays appear as well defined peaks in the $M(\mu^+\mu^-)$ and $M(e^+e^-)$ distributions shown in Figs.~1(a,b).  The event distributions were fit with constant backgrounds and Gaussian peak shapes.  For $J/\psi\rightarrow\mu^+\mu^-$ decays we obtain $M(J/\psi) = 3096.5 \pm 0.5$~MeV,  $N_\mu=137\pm15$~counts  and a fitted resolution width, $\sigma_\mu=4.2\pm0.5$~MeV. For $J/\psi\rightarrow e^+e^-$ decays we obtain $M(J/\psi) = 3096.3 \pm 0.5$~MeV, $N_e=96\pm11$~counts and a fitted width,  $\sigma_e=3.9 \pm 0.5$~MeV. Monte Carlo (MC) determined  efficiencies for decays containing $\mu^\pm$  are $\epsilon_\mu=53\%$, and for decays containing $e^\pm$  are $\epsilon_e=43\%$.  
%We estimate the radiative correction factor $C=0.82$. 
To obtain Born cross sections, we correct the observed cross sections for the effect of initial state radiation (ISR) using the method of Bonneau and Martin \cite{bonneaumartin}. 
The Born cross sections are determined as $\sigma_{\mathrm{Born}}(e^+e^- \to \pi^+\pi^-J/\psi) = N / \mathcal{L}\,\epsilon\,C\,\mathcal{B}$, where $\mathcal{B}\equiv\mathcal{B}(J/\psi\to e^+e^-,\mu^+\mu^-)=5.9\%$~\cite{pdg}, and the radiation correction factor $C=0.82$. The cross sections are $\sigma_{\mathrm{Born}}(\mu^\pm) = 9.1 \pm 1.0$(stat)~pb, and $\sigma_{\mathrm{Born}}(e^\pm) = 7.9\pm 0.9$(stat)~pb, with the average value of $\sigma_{\mathrm{Born}}(e^+e^-\rightarrow\pi^+\pi^-J/\psi)=8.4\pm0.7$(stat)~pb. This is nearly 1/7~th of the cross section observed by BES~III in decays of Y(4260).

\begin{table*}[!tb]
\caption{Summary of results for $\psi(4160)\rightarrow\pi^+\pi^-J/\psi$ and $\pi^0\pi^0J/\psi$. The branching fractions, 
$\mathcal{B}(J/\psi\rightarrow e^+e^-)=\mathcal{B}(J/\psi\rightarrow \mu^+\mu^-)=5.9\%$, and the ISR correction factor $C=0.82$ are used to obtain 
$\sigma_\mathrm{Born}= N / \mathcal{L}\,\epsilon\,C\,\mathcal{B}$. 
The cross sections $\sigma_{\mu^+\mu^-}$ and $\sigma_{e^+e^-}$ 
refer to $J/\psi$ decays to 
$\mu^+\mu^-$ and $e^+e^-$, respectively.
The errors on all quantities are statistical only.
Systematic errors on $\sigma(e^+e^- \to \pi\pi J/\psi)$ are estimated to be 5\%.}

\setlength{\tabcolsep}{2pt}

\begin{center}
\begin{tabular}{llccccccccc}
\hline\hline
 & & $\sqrt{s}$ & $\mathcal{L}$ & $\epsilon_{\mu}/\epsilon_e{}$& $N$ & $N$  & $\sigma_{\mu^+\mu^-}$ & $\sigma_{e^+e^-}$ & $\langle \sigma(e^+e^- \to \pi\pi J/\psi) \rangle$  \\
 & & GeV & pb$^{-1}$ & & $(J/\psi\to\mu^+\mu^-)$ & $(J/\psi\to e^+e^-)$ & pb& pb & pb  \\
\hline
BES III~\cite{besz} & $\pi^+\pi^-$ & 4.26 & 525 & 0.54/0.38& $882\pm33$ & $595\pm28$   & $64.4\pm2.4$ & $60.7\pm2.9$ & $62.9\pm1.9$   \\[5pt]
%%Belle & 4.26 & 96700 & --- & --- & --- & --- & $5.2$ & $3895\pm8$ & $63\m35$ & $21\pm3$ \\

Present & $\pi^+\pi^-$ & 4.17 & 586 & 0.53/0.43& $137\pm15$ & $96\pm11$ & $9.1\pm1.0$ & $7.9\pm0.9$ &  $8.4\pm0.7$  \\
        & $\pi^0\pi^0$ & 4.17 & 586 & 0.23/0.18&  $40\pm 8$ & $29\pm 5$ & $6.1\pm1.2$ & $5.7\pm1.0$ &  $5.9\pm0.8$  \\
\hline\hline
\end{tabular}
\end{center}
\end{table*}

\begin{table*}[!tb]
\caption{Summary of fit results for $Z_c^{\pm}$ and $Z_c^0$. The second uncertainties in $M(Z_c)$, $\Gamma(Z_c)$ and $R$ are systematic.  The log-likelihood determined significance levels for the present results have been calculated including systematic errors.}

\setlength{\tabcolsep}{2pt}

\begin{center}
\begin{tabular}{llcccccc}
\hline\hline
 & & Significance & $N(Z_c)$& $\langle \sigma(e^+e^-\rightarrow\pi Z_c\rightarrow\pi\pi J/\psi) \rangle$, pb  & $M(Z_c)$, MeV & $\Gamma(Z_c)$, MeV &  $R$, $\%$ \\
% & & GeV & pb$^{-1}$ & \multicolumn{1}{r}{$\mu^+\mu^-/e^+e^-)$} & \multicolumn{1}{r}{$\pi\pi J/\psi)$,~pb} &  & $\sigma$ & MeV & MeV & \% \\
\hline
BES III~\cite{besz} & $\pi^+\pi^-$   & $>8\sigma$ & $307\pm48$ & $13.5\pm2.1$ & $3899\pm4\pm5$ & $46\pm10\pm20$ & $22\pm3\pm8~$ \\
Belle~\cite{bellez} & $\pi^+\pi^-$   & $>5.2\sigma$ & $159\pm49$ & --- & $3895\pm7\pm5$ & $63\pm24\pm26$ & $29\pm9$ \\[5pt]

Present: \\
~~($M^2(\pi^+\pi^-)$ --- all) & $\pi^+\pi^-$   & $5.1\sigma$& $71\pm15$ & $2.7\pm0.6\pm0.8$ & $3886\pm4\pm 2$ & $33\pm6\pm7$ & $32\pm8\pm 10$ \\
~~($M^2(\pi^+\pi^-)<0.65$~GeV) & $\pi^+\pi^-$ & $5.7\sigma$& $81\pm16$ & $3.9\pm0.8\pm0.8$ & $3886\pm4\pm 2$ & $37\pm4\pm8$ & $52\pm12\pm 10$ \\
~~($M^2(\pi^0\pi^0)<0.65$~GeV) & $\pi^0\pi^0$ & $3.5\sigma$& $25\pm7$  & $3.2\pm0.8\pm0.6$ & $3904\pm9\pm 5$ & $37$ (fixed)  & $63\pm21\pm 11$ \\
\hline\hline
\end{tabular}
\end{center}
\end{table*}

%We define the  $J/\psi$ resonance region as $M(e^+e^-,\mu^+\mu^-)=3097\pm20?$~MeV, and the sideband regions as $M(e^+e^-,\mu^+\mu^-)=3030\pm30$~MeV and $M(e^+e^-,\mu^+\mu^-)=3170\pm30$~MeV. We construct distributions of $M(\pi^+J/\psi)$ and $M(\pi^-J/\psi)$,  and the maximum of the two, labeled  $M_\text{max}(\pi^\pm J/\psi)$, the same observable as BES~III and Belle, in the two regions.
%Both mass distributions  $M(\pi^+J/\psi)$ and $M(\pi^-J/\psi)$ show enhancements over background in the vicinity of 3900~MeV, but as observed by BES~III and Belle, the background is much reduced in the $M_\text{max}(\pi^\pm J/\psi)$ distribution with the removal of the wrong combinations.

Fig.~1(c) shows the Dalitz plot distribution of $M^2(\pi^+\pi^-)$ versus $M^2(\pi^+J/\psi)$.  A clear enhancement of events is seen near $M^2(\pi^+J/\psi)\approx15$~GeV$^2$, and its reflection at $\approx11$~GeV$^2$.  The enhancement at $\sim15$~GeV$^2$ appears to be confined mostly to the $M^2(\pi^+\pi^-)<0.65$~GeV$^2$ region.
Fig.~1(d) shows the $M(\pi^+\pi^-)$ distribution corresponding to the $M^2(\pi^+\pi^-)$ projection of the Dalitz plot of Fig.~1(c).  We note that our $M(\pi^+\pi^-)$ distribution arising from $\psi(4160)\to\pi^+\pi^-J/\psi$ is quite different from the distributions observed by BES~III and Belle for $\text{Y}(4260) \to \pi^+\pi^- J/\psi$.  Our $M(\pi^+\pi^-)$ distribution is nearly flat with a small enhancement corresponding to $f_0(980)$, whereas the $M(\pi^+\pi^-)$ distributions observed by both BES~III and Belle show a large enhancement due to $f_0(980)$ and a deep minimum at $M(\pi^+\pi^-)\approx550$~MeV.  This difference indicates that our observations are characteristic of the decay of $\psi(4160)$, and are not due to any substantial contribution from the tail of Y(4260).

We consider the maximum of $M(\pi^+J/\psi)$ and $M(\pi^-J/\psi)$, defined as $M_\text{max}(\pi^\pm J/\psi)$.  Its distribution for the full range of $M^2(\pi^+\pi^-)$ is shown in Fig.~2(a).
As noted before, the enhancement in the Dalitz plot of Fig.~1(c) appears to be confined to the $M^2(\pi^+\pi^-)<0.65$~GeV$^2$ region.  %The projection of $M(\pi^+\pi^-)$ reflects this in the much larger yield in the $M(\pi^+\pi^-)<0.8$~GeV region.  
This suggests that we should divide our data in two parts, and study $M_\text{max}(\pi^\pm J/\psi)$ distributions separately in the two regions, $M^2(\pi^+\pi^-)<0.65$~GeV$^2$ and $>0.65$~GeV$^2$.  The distributions of $M_\text{max}(\pi^\pm J/\psi)$ corresponding to the two $M^2(\pi^+\pi^-)$ regions are shown in Fig.~2(b) and 2(c).

Both Fig.~2(a), with no $M^2(\pi^+\pi^-)$ cut, and Fig.~2(b), for $M^2(\pi^+\pi^-)<0.65$~GeV$^2$, show a clear peak at $M(\pi^\pm J/\psi)\approx3900$~MeV. In contrast, Fig.~2(c) for $M^2(\pi^+\pi^-)>0.65$~GeV$^2$ shows no evidence for it.  The $M^2(\pi^+\pi^-)<0.65$~GeV$^2$ distribution has a factor two smaller background and no loss of signal counts compared to that for the whole distribution. It leads to a better resolution of the peak at $M_\text{max}(\pi^\pm J/\psi)\approx3900$~MeV from the background.  %We therefore derive our results for $Z_c^\pm$ by fitting the distribution in Fig.~2(a) for $M^2(\pi^+\pi^-)<0.65$~GeV$^2$.

The event distributions in Figs.~2(a,b) are fitted with a background and a peak.  We use Monte Carlo-determined phase space shapes for the backgrounds.
As shown in the figures, the $M_{\mathrm{max}}(\pi^{\pm}J/\psi)$ distribution for the $J/\psi$ sidebands has a shape which is similar to the shape of the phase space distribution, but is only about half of the observed background in Figs.~2(a) and 2(b).
The peak shape was obtained by convolving the MC determined instrumental width of 3.6~MeV with a Breit-Wigner resonance. The normalization of the background, the position of the peak, its width, and its magnitude were kept free in the fit. The fits are binned likelihood fits with no interference assumed between the resonance and the background.  
The results of the fits in Figs.~2(a) and 2(b) are listed separately in Table~II.
%The results of the fits are: Fig.~2(a) for all $M^2(\pi^+\pi^-)$ events: $N=71\pm15$, $M(Z^{\pm}_c) = 3886\pm4(\text{stat})$~MeV, $\Gamma(Z^{\pm}_c) = 33\pm6(\text{stat})$~MeV, $\chi^2/d.o.f. = 33/38$; Fig.~2(b) for $M^2(\pi^+\pi^-)<0.65$~GeV$^2$: $N= 81\pm16$, $M(Z^{\pm}_c) = 3886\pm4(\text{stat})$~MeV, $\Gamma(Z^{\pm}_c) = 37\pm4(\text{stat})$~MeV, and $\chi^2/d.o.f.=33/35$.
%We use the standard method to determine the significance of the peak as $(\sqrt{-2\ln(L_0/L_{max})})\sigma$, where $L_{max}$ is the maximum likelihood returned by the fit including the peak, and $L_0$ is the likelihood returned by the fit without the peak.
For both $M(Z_c^\pm)$ and $\Gamma(Z_c^\pm)$ our results agree with those of BES~III and Belle.  We determine the significance of the peaks by using the difference in likelihoods between fits with and without the peaks, taking into account the difference in degrees of freedom between the two fits.
The significance of the peak for 3 degrees of freedom is $5.4\sigma$ for all $M^2(\pi^+\pi^-)$ events in Fig.~2(a), and $6.0\sigma$ for $M^2(\pi^+\pi^-)<0.65$~GeV$^2$ events in Fig.~2(b).

\begin{figure}[!tb]

\begin{center}
\includegraphics[width=2.4in]{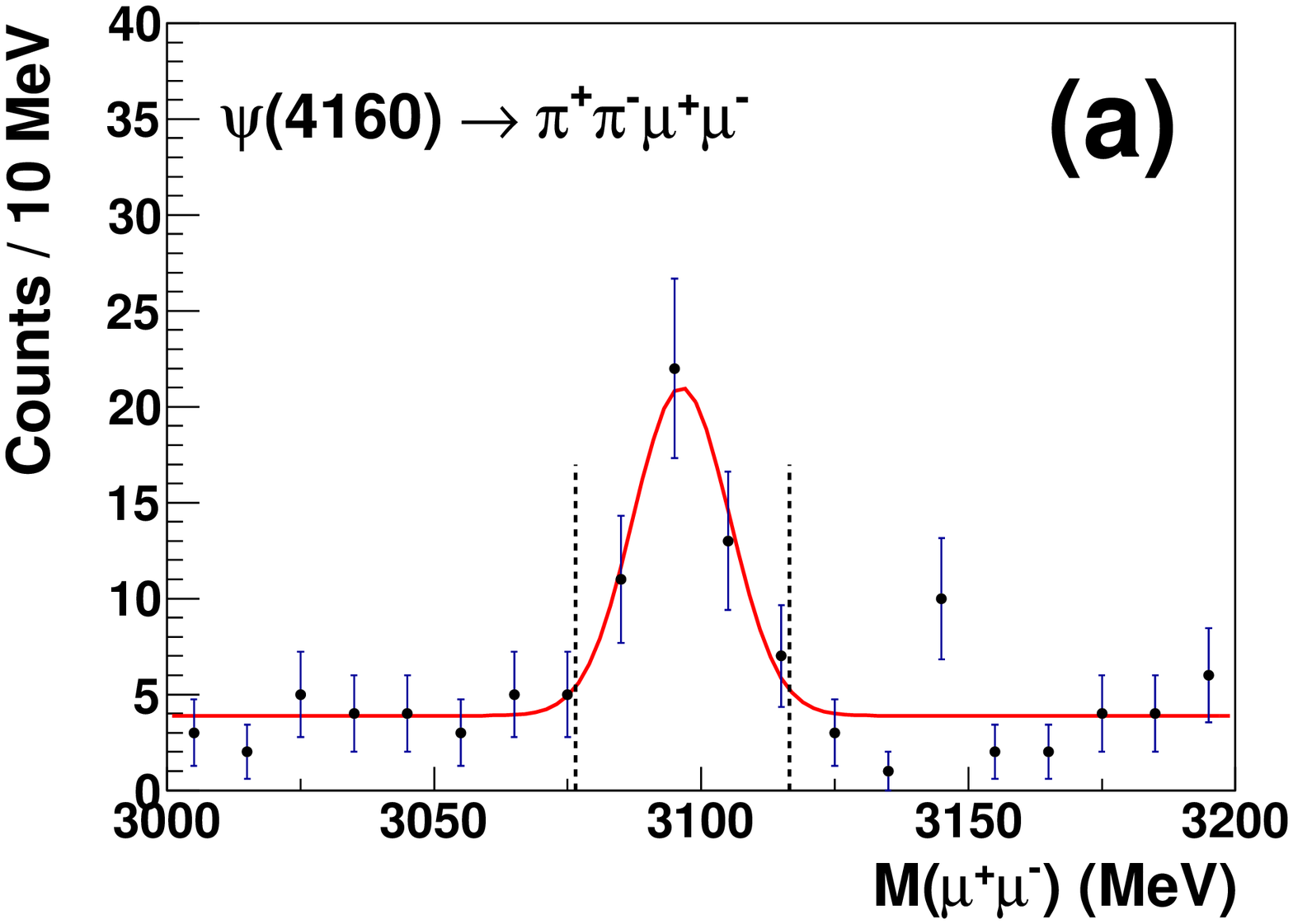}

\includegraphics[width=2.4in]{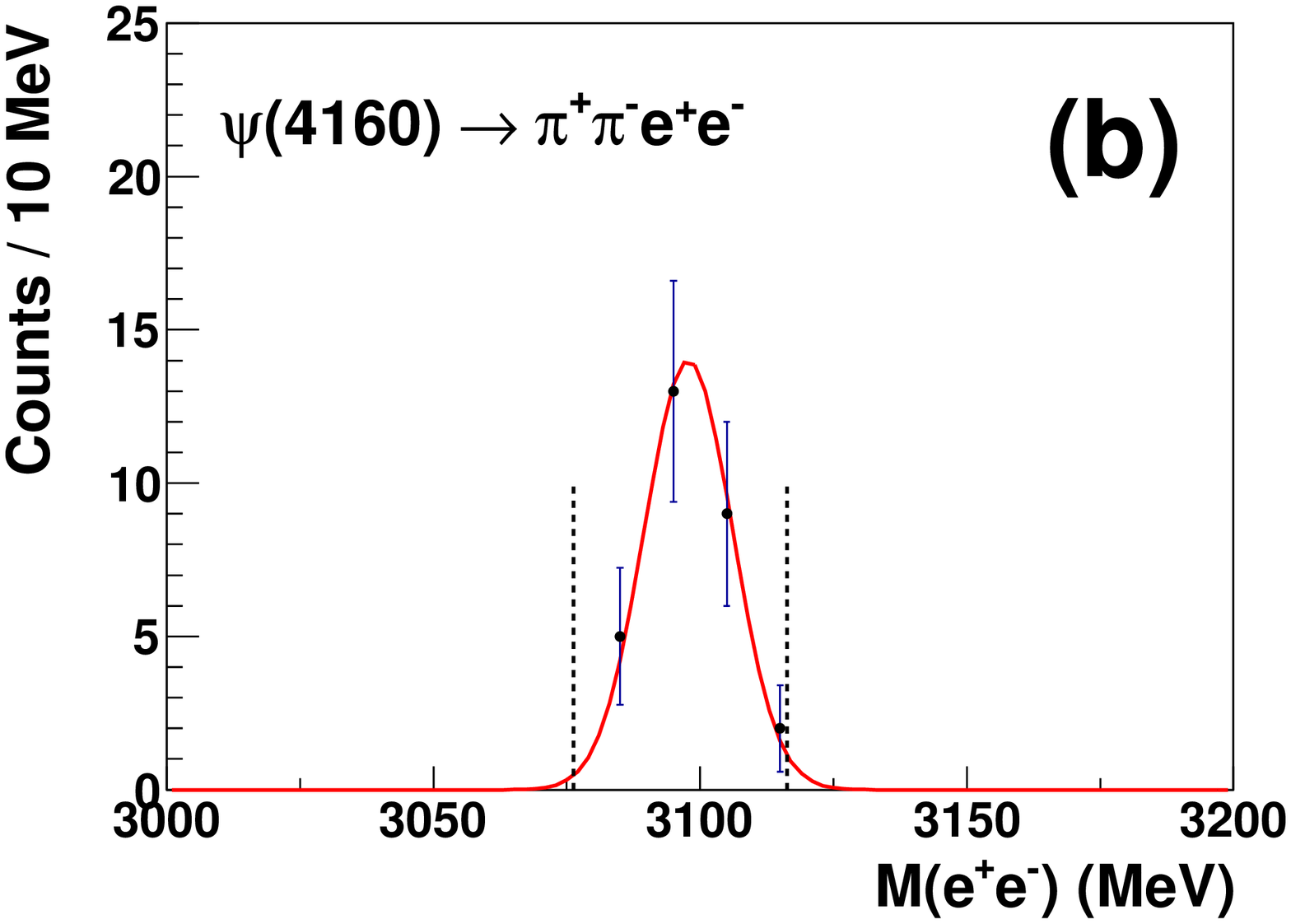}
%\includegraphics[width=2.4in]{neut-ee_jpsi.eps}
%\includegraphics[width=2.4in]{neut/jpsi/ee_jpsi.eps}

%\vspace*{10pt}

%\vspace*{10pt}

\includegraphics[width=2.4in]{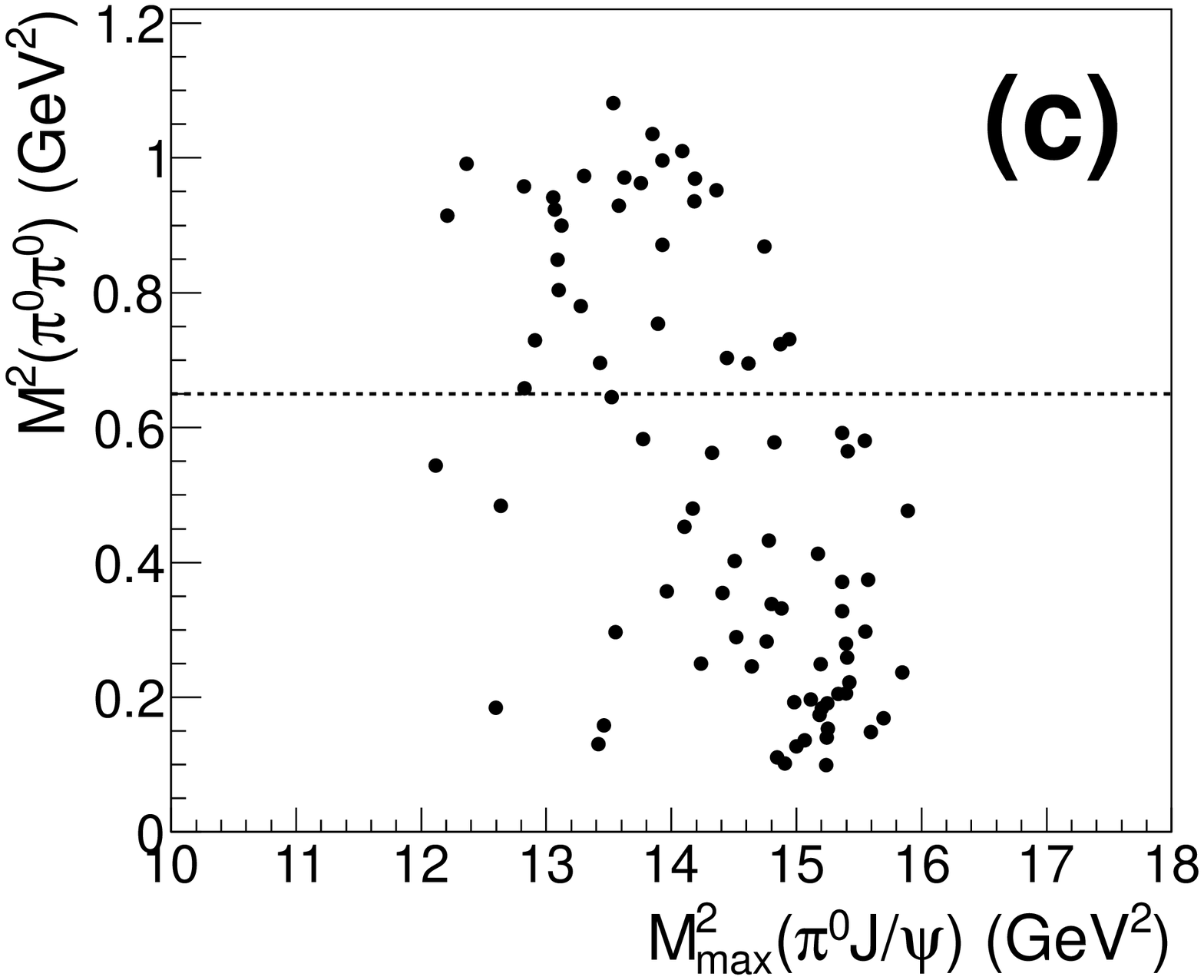}
%\includegraphics[width=2.4in]{neut-2Dleft.eps}
%\includegraphics[width=2.4in]{neut/2Dleft/2Dleft.eps}

%\vspace*{10pt}

\includegraphics[width=2.4in]{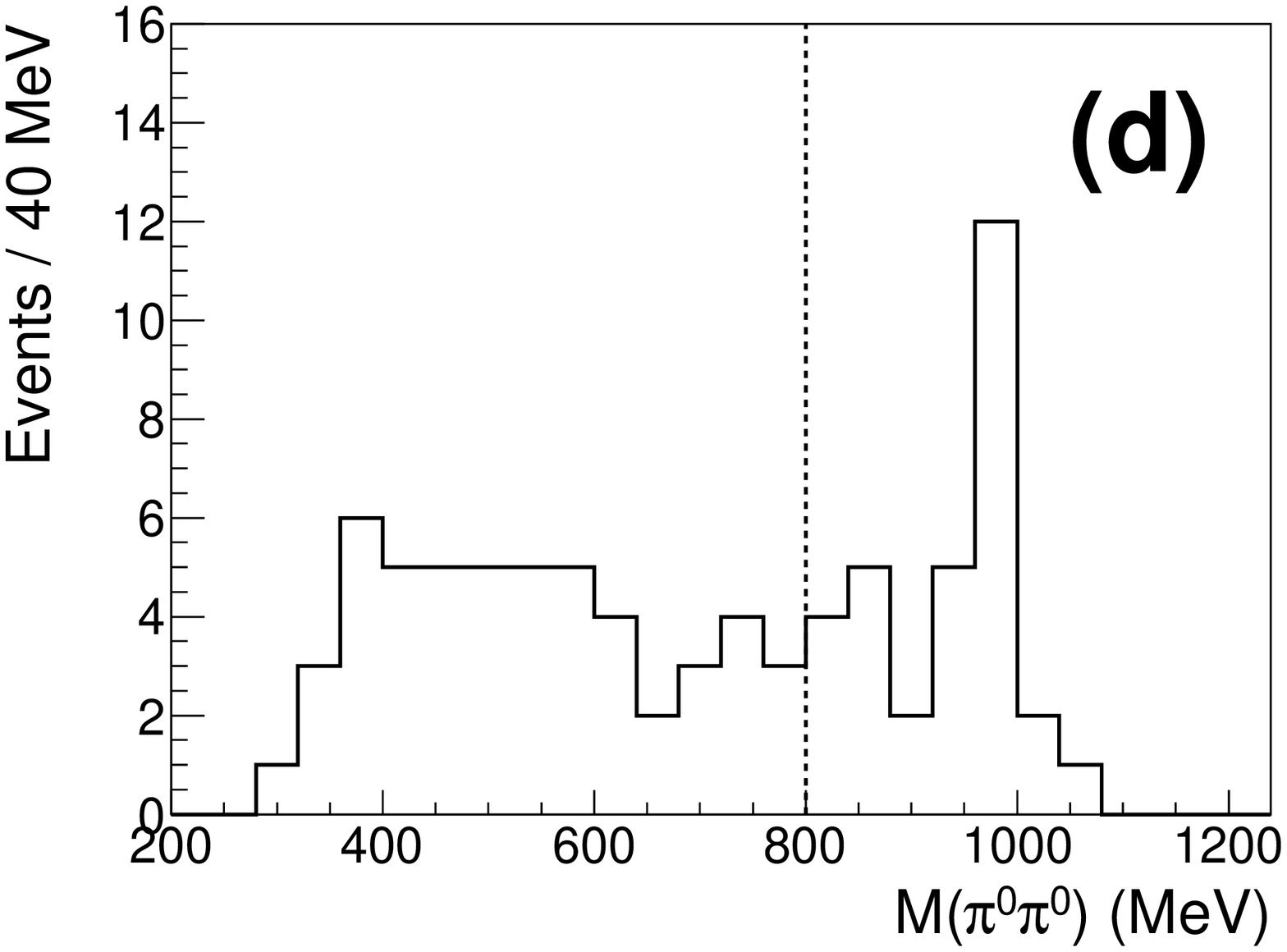}
\end{center}

\caption{(a,b) Invariant mass distributions for $\psi(4160)\rightarrow\pi^0\pi^0(\mu^+\mu^-,e^+e^-)$. The curves are fits as described in the text. The dashed vertical lines define the $J/\psi$ resonance region used in subsequent analysis; (c) Dalitz plot for $M^2(\pi^0\pi^0)$ versus $M^2(\pi^0J/\psi)$; (d) $M(\pi^0\pi^0)$ projection of the Dalitz plot.  The dashed lines in (c) and (d) indicate division of the data into two parts, as described in the text.}

\end{figure}

\begin{figure}[!t]

\begin{center}
\includegraphics[width=2.8in]{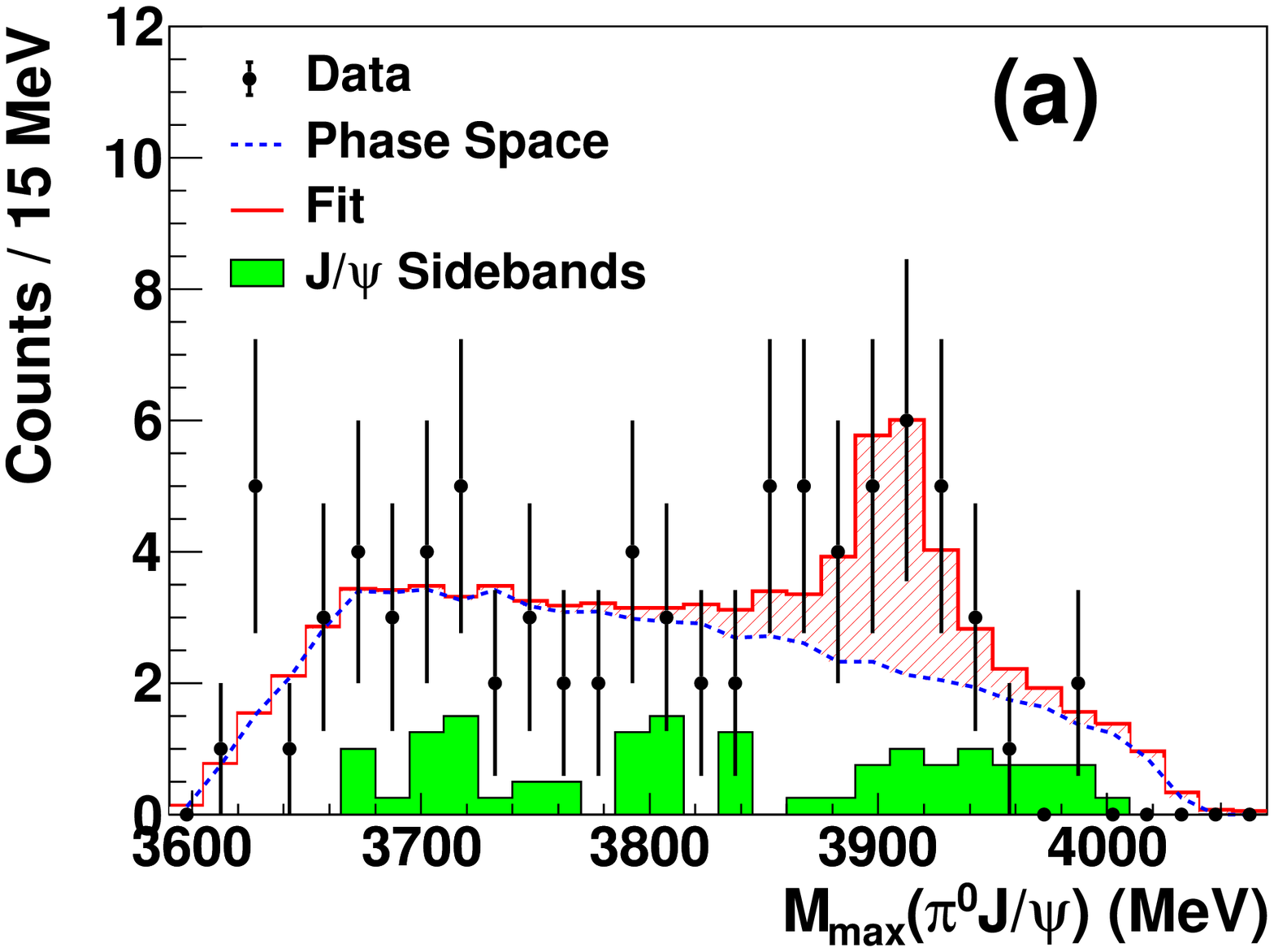}

\includegraphics[width=2.8in]{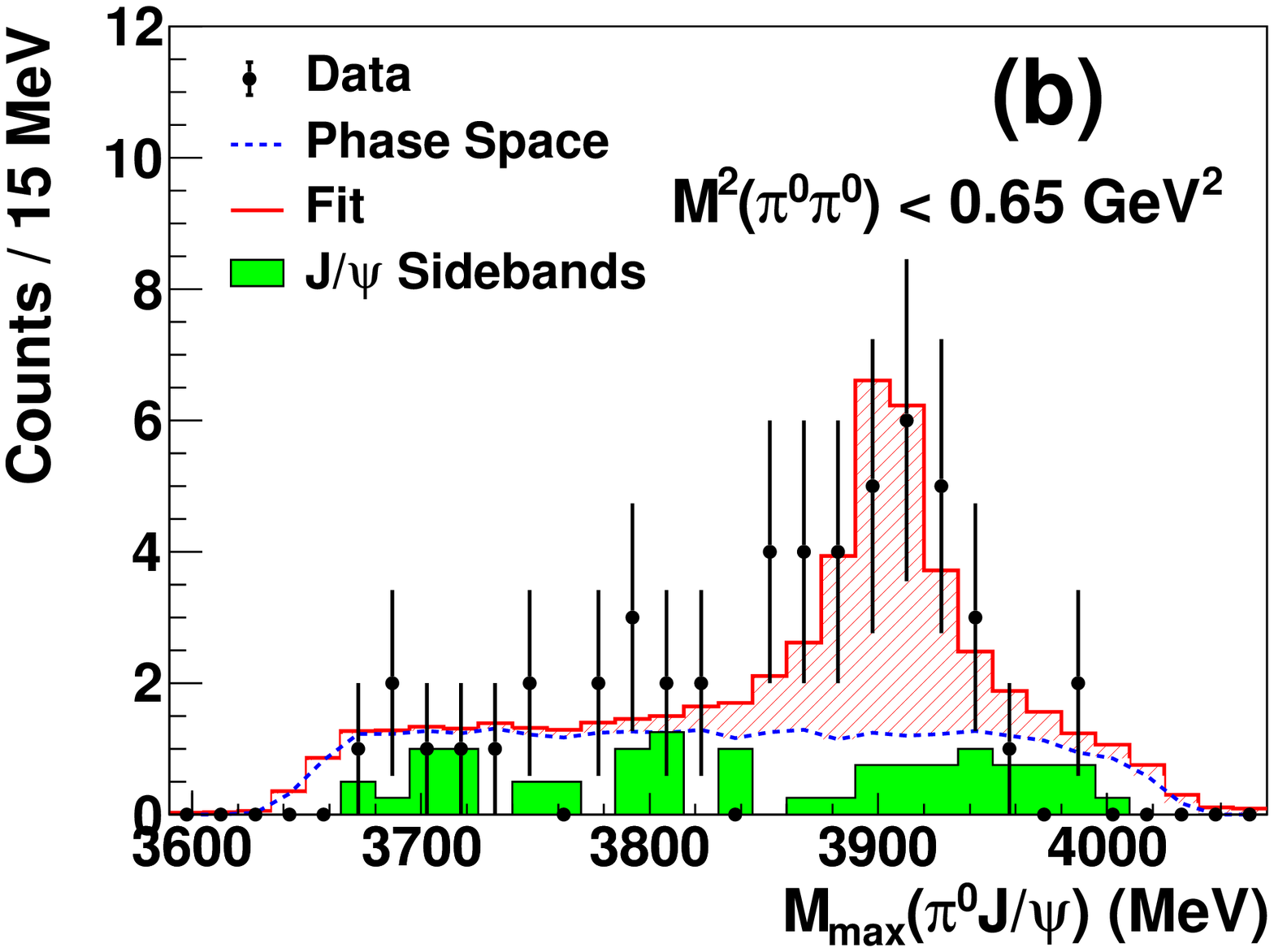}
%\includegraphics[width=2.8in]{neut-max_mass.eps}
%\includegraphics[width=2.8in]{neut-pi0jpsi.eps}
%\includegraphics[width=2.8in]{neut/pipi08bin15_pi0jpsi/pi0jpsi.eps}
%\includegraphics[width=2.8in]{neut/pipi08bin15_pi0jpsi-new/max2_mass.eps}

%\vspace*{10pt}

%\includegraphics[width=2.8in]{neut-bigpipi_jpsipi.eps}
%\includegraphics[width=2.8in]{neut/pipi08_bigpipi/bigpipi_jpsipi.eps}
%\includegraphics[width=2.8in]{neut/bigpipi-new/sideband_bigpipi.eps}
%\includegraphics[width=2.8in]{neut-sideband_bigpipi.eps}
\includegraphics[width=2.8in]{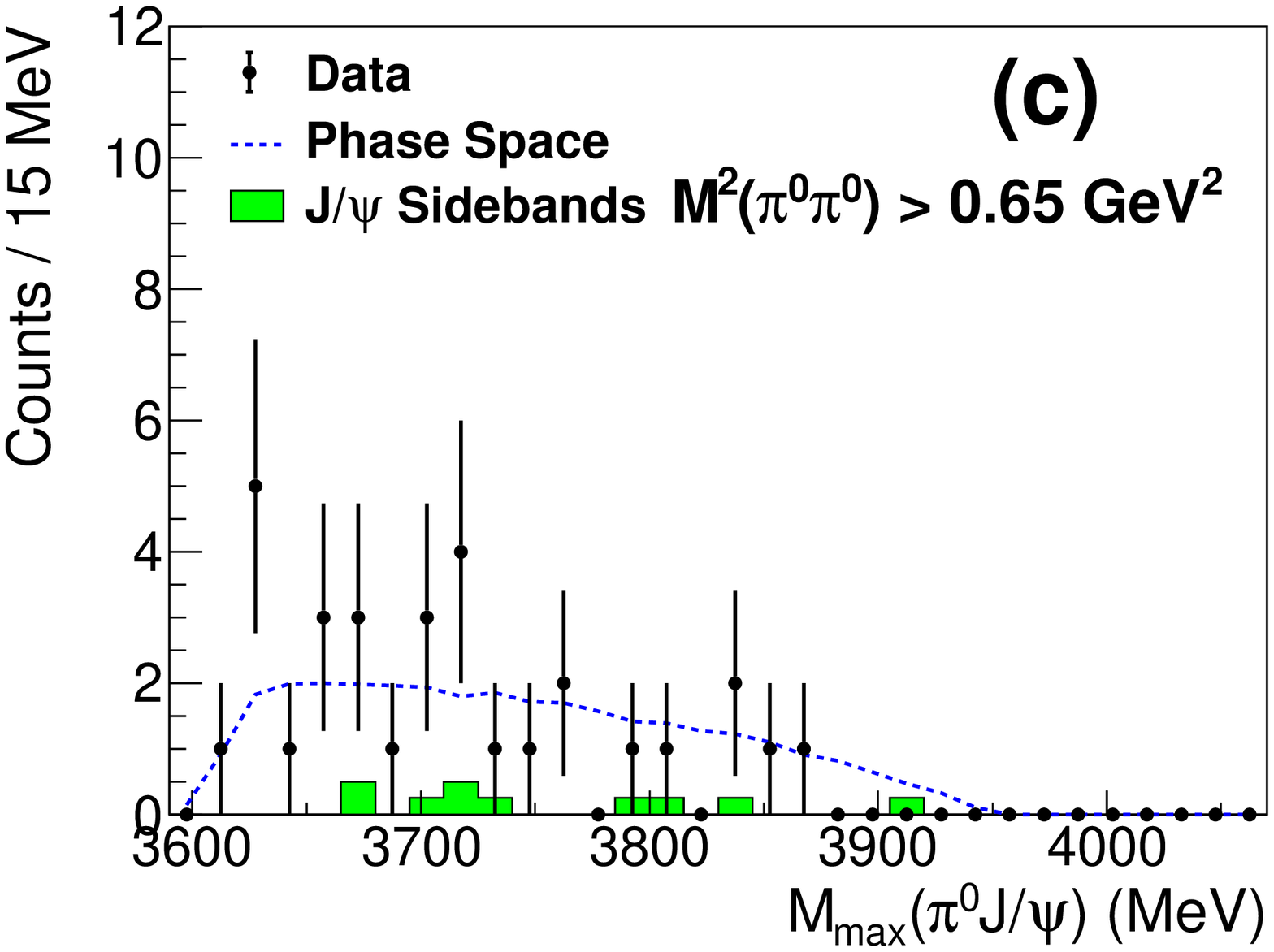}
\end{center}

\caption{Distributions of $M_\text{max}(\pi^0 J/\psi)$ as observed in the decay $\psi(4160)\rightarrow\pi^0\pi^0J/\psi$. The histograms show the fits.  The dashed curves show the MC-determined phase-space background.  The hatched peak shows the contribution of the Breit-Wigner resonance. (a) with no cut in $M^2(\pi^0\pi^0)$, (b) $M^2(\pi^0\pi^0)<0.65$~GeV$^2$, (c) $M^2(\pi^0\pi^0)>0.65$~GeV$^2$.}

\end{figure}

The $Z_c^\pm$ state decaying into $J/\psi\pi^\pm$ has isospin $I=1$.  The identification of the neutral member of this ``exotic'' isospin triplet has not been reported by either BES~III or Belle.  We report here the evidence for $Z_c^0$ in the decay $\psi(4160)\rightarrow\pi^0\pi^0J/\psi$.  Taking account of isospin and the nearly factor two smaller efficiency for detecting $\pi^0\pi^0$ compared to $\pi^+\pi^-$, we expect to observe $\sim60~J/\psi\pi^0\pi^0$ events, and $\sim20$ $Z_c^0$.  Although the expected statistics are limited, we have made this measurement.  

The events for $\psi(4160)\rightarrow\pi^0\pi^0J/\psi$, $\pi^0\rightarrow2\gamma$, $J/\psi\rightarrow\mu^+\mu^-,e^+e^-$ are required to have only two oppositely charged particles and at least four photons.  Photon candidates are identified as showers in the calorimeter which are not associated with the projection of any charged track, have a transverse energy spread consistent with that of an electromagnetic shower, and have a total energy of $>30$~MeV. The $\pi^0\to\gamma\gamma$ candidates are required to have a mass consistent with $M(\pi^0)$ within $3\sigma$.  The $\pi^0$ candidates are then kinematically fitted by constraining the mass of the photon pair to the known $M(\pi^0)$.  Each event is required to have at least two $\pi^0$ candidates.  The 4C kinematic fit is performed on all $\pi^0\pi^0J/\psi$ candidates in the event, and the one with the smallest $\chi^2$ is retained.

\begin{table*}

\caption{Summary of systematic uncertainties.}

\begin{center}
\begin{tabular}{lccccc}
\hline
sources of & \multicolumn{3}{c}{ $\pi^+\pi^-$ ($M^2(\pi^+\pi^-)\,-\,\text{all}\,/<0.65$) } & \multicolumn{2}{c}{ $\pi^0\pi^0$ ($M^2(\pi^0\pi^0)< 0.65$), $\Gamma$~fixed } \\
systematic errors & $\Delta M$ (MeV) & $\Delta \Gamma$ (MeV) & $\Delta \sigma$ (\%) & $\Delta M$ (MeV) & $\Delta \sigma$ (\%) \\
\hline
Mass calibration & 0.5 & --- & --- & 4.0 & --- \\
Background$^*$ & $1.0/1.3$ & $5.1/5.8$ & $28/17$ & 0.4 & 4 \\
BW parameterization$^\dagger$ & $1.7/1.0$ & $0.9/1.8$ & $2/3$ & 3.0 & 2 \\
Instrumental width ($\pm10\%$) & $<0.1$ & $<0.4$ & $<1$ & 0.1 & 1 \\
Bin size ($7.5-12.5$~MeV) & $1.0/0.5$ & $3.4/3.6$ & $6/4$ & 1.6 & 3 \\
Fit range (in $3.60-4.05$~GeV) & $0.2/0.4$ & $1.9/2.4$ & $5/5$ & 0.4 & 4 \\
$\chi^2_\text{4C~fit}$ & $0.9/0.8$ & $1.5/1.9$ & $10/7$ & 1.6 & 6 \\
$\mathcal{L},\mathcal{B},C$, event reconstruction & --- & --- & 5 & --- & 10 \\
Fixed width variation & --- & --- & --- & 0.8 & 12 \\
\hline
Sum in quadrature & 2.4/2.0 & 6.7/7.7 & 31/20 & 5.4 & 18  \\
\hline
\end{tabular}
\end{center}
$^*$ Phase space versus phase space$\times$polynomial.\\
$^\dagger$ Mass-dependent width, phase space terms, and spin-1 versus spin-0.

\end{table*}

As shown in Fig.~3(a,b), we again see clear $J/\psi$ peaks in the $M(\mu^+\mu^-,e^+e^-)$ distributions. The $\pi^0\pi^0\mu^+\mu^-$ spectrum in Fig.~3(a) has a small number of background events, and the $\pi^0\pi^0e^+e^-$ spectrum in Fig.~3(b) has none.  Monte Carlo simulations show that in absence of other charged particles, rejection of electron background in the $\pi^0\pi^0e^+e^-$ spectrum is more than an order of magnitude more effective than in the $\pi^+\pi^-e^+e^-$ spectrum of Fig.~1(b). This accounts for the absence of background counts in the $\pi^0\pi^0e^+e^-$ spectrum of Fig.~3(b).
Fits to the spectra yield for the $\mu^+\mu^-$ case $M(J/\psi)=3096.3\pm2.1$~MeV, $N_\mu=40\pm8$, and a resolution width of $\sigma_\mu=9.0\pm2.0$~MeV, and for the $e^+e^-$ case $M(J/\psi)=3097.8\pm1.5$~MeV, $N_e=29\pm5$, and a resolution width of $\sigma_e=8.3\pm1.1$~MeV. 
The MC determined efficiency for decays containing $\mu^\pm$ is $\epsilon_\mu=23\%$ and for decays containing $e^\pm$ is $\epsilon_e=18\%$.  The cross sections $\sigma_{\mathrm{Born}}(\pi^0\pi^0J/\psi)$, determined in the same manner as for the charged pion case are $\sigma_{\mathrm{Born}}(\mu^\pm)=6.1\pm1.2$(stat)~pb and $\sigma_{\mathrm{Born}}(e^\pm)=5.7\pm1.0$(stat)~pb with the average $\sigma_{\mathrm{Born}}(e^+e^-\rightarrow\pi^0\pi^0 J/\psi)=5.9\pm0.8$(stat)~pb.  
Fig.~3(c) shows the Dalitz plot distribution of $M^2(\pi^0\pi^0)$ versus $M^2(\pi^0J/\psi)$, and
Fig.~3(d) shows the $M(\pi^0\pi^0)$ distribution corresponding to the $M^2(\pi^0\pi^0)$ Dalitz plot projection.
%Despite the small statistics, a clear enhancement of events is seen near $M^2(\pi^+J/\psi)\approx15$~GeV$^2$, and its reflection at $\approx11$~GeV$^2$.  

The observed distributions of $M_\text{max}(\pi^0 J/\psi)$ events in the two regions, $M^2(\pi^0\pi^0)<0.65$~GeV$^2$ and$M^2(\pi^0\pi^0)>0.65$~GeV$^2$, are shown in Figs.~4(a,b).  
The peak at $M_\text{max}(\pi^0 J/\psi)\approx3900$~MeV is seen in Fig.~4(a) for events with all $M^2(\pi^0\pi^0)$, but it is much more clearly delineated in Fig.~4(b) for $M^2(\pi^0\pi^0)<0.65$~GeV$^2$ because of a nearly factor two smaller background.  Fig.~4(c) shows that there is no evidence for a peak anywhere for $M^2(\pi^0\pi^0)>0.65$~GeV$^2$.
We fit the $M_\text{max}(\pi^0 J/\psi)$ distribution in Fig.~4(b) in exactly the same manner as the $M_\text{max}(\pi^\pm J/\psi)$ distributions in Figs.~2(a) and 2(b), with phase space background, and peak shape obtained by convolving the MC determined instrumental peak shape, with width of $5.4$ MeV, with a Breit-Wigner function.  
For $M(Z_c^0)$, $\Gamma(Z_c^0)$, and $N(Z_c^0)$ all kept free, the result is $M(Z_c^0)=3901\pm4(\text{stat})$~MeV, $\Gamma(Z_c^0)=58\pm27(\text{stat})$~MeV, and $N(Z_c^0)=31\pm10$, and $\chi^2/d.o.f.=13/20$.  The large values of the width and its error indicate that with the small event statistics it is not advisible to keep all three parameters free.  We therefore fix the width $\Gamma(Z_c^0)=37$~MeV, as determined for $Z_c^\pm$, and include its uncertainty of $\pm13$~MeV as a source of systematic error.  For the corresponding fit, shown in Fig.~4(b), the results are $N(Z_c^0)=25\pm7$~counts, $M(Z_c^0)=3904\pm9$(stat)~MeV, $\Gamma(Z_c^0)=37$(fixed)~MeV, as also listed in Table~II.  The $\chi^2/d.o.f.$ of the fit is $14/21$.  The likelihood-determined significance of the $Z^0_c$ peak is $3.7\sigma$ for 2 degrees of freedom.

We have evaluated systematic uncertainties due to many sources.
The sources which only contribute systematic uncertainties in cross sections are: luminosity~(1\%), $\mathcal{B}(J/\psi\to l^+l^-)$~(1\%), radiative correction $C$~(2\%), and event reconstruction, $l^+l^-\pi^+\pi^-$ ($4\times1\%$), $l^+l^-\pi^0\pi^0$ ($2\times1\% + 2\times5\%$).  In quadrature, these add to 5\% for $J/\psi\pi^+\pi^-$ and 10\% for $J/\psi\pi^0\pi^0$.  Additional sources and their contributions to systematic uncertainties are listed in Table~III for (all $M(\pi^+\pi^-)/(M(\pi^+\pi^-)<0.65~\text{GeV}^2)$) and for $M(\pi^0\pi^0)<0.65$~GeV$^2$.
Inclusion of these systematic uncertainties reduce the significance of the $Z_c^\pm$ peak from $5.4\sigma$ to $5.1\sigma$ for all $M^2(\pi^+\pi^-)$, from $6.0\sigma$ to $5.7\sigma$ for $M^2(\pi^+\pi^-)<0.65$~GeV$^2$, and $3.7\sigma$ to $3.5\sigma$ for the $Z_c^0$ peak.

%Our final results for $\psi(4160)\rightarrow\pi^0\pi^0J/\psi$ and $Z^0_c$ are also presented in Tables I and II. The results for $Z^0_c$ are in agreement with those for $Z_c^\pm$ within their statistical errors.  

We summarize our final results for $\psi(4160)\rightarrow(\pi^+\pi^-,\pi^0\pi^0)J/\psi$ and $Z^{\pm,0}_c$ in Tables I and II. The results for $Z^0_c$ are in agreement with those for $Z_c^\pm$ within their statistical errors.  

As seen in the Table~II our results for $M(Z_c^\pm)$ are in good agreement with those reported by BES~III~\cite{besz} and Belle~\cite{bellez}.  Our results for $\Gamma(Z_c^\pm)$ are smaller, but are in agreement with the BES~III and Belle results which have large errors. 

The cross section ratio is defined as
$$R(Z^{\pm}_c) \equiv \frac{\sigma(e^+e^- \to \pi^\pm Z_c^\mp(3900) \to \pi^+\pi^-J/\psi)}{\sigma(e^+e^- \to \pi^+\pi^-J/\psi)}.$$
For all $M^2(\pi^+\pi^-)$ events, we obtain $R(Z_c^\pm)=(32\pm8\pm10)\%$, in agreement with $R(Z_c^\pm) = (22\pm3\pm8)\%$ reported by BES~III~\cite{besz}, and $(29\pm9)\%$ reported by Belle~\cite{bellez}.
For events with $M^2(\pi^+\pi^-)<0.65$~GeV$^2$, $\sigma(e^+e^-\to\pi^+\pi^- J/\psi)= 7.5\pm0.8$~pb, and $ R(Z^{\pm}_c)= (52\pm12\pm 10)\%$, which is substantially larger.  The reason for this difference is that in both BES~II and Belle measurements, the cross section, $\sigma(e^+e^-\to\pi^+\pi^-J/\psi)$ in the denominator of the expression for $R(Z_c^\pm)$ contains large contribution from $f_0(980)$, which lead to a small value of $R$.  In our case, this contibution is small, and for $M^2(\pi^+\pi^-)<0.65$~GeV$^2$, almost zero.
$R(Z_c^0)=(63\pm21\pm11)\%$ agrees with $R(Z_c^\pm)=(52\pm12\pm10)\%$ for $M^2(\pi^+\pi^-)<0.65$~GeV$^2$.

To summarize, our observations provide independent confirmation of the existence of the $Z_c^\pm(3900)$ state, and provide new evidence for the existence of the neutral member $Z_c^0(3900)$ of this isospin triplet.  Our observations of $Z_c^{\pm, 0}(3900)$ are based on $e^+e^-$ annihilations at $\sqrt{s}=4170$~MeV, on the peak of the well-established $2^3D_1$ vector state $\psi(4160)$ of charmonium, and are distinct from the observations of BES~III and Belle made in the decays of Y(4260).
  %%, and we do not need to attribute any unconventional properties to the initial state, as may be invoked for the observation of $Z_c^\pm$ at Y(4260).
%However, we note that since both $\psi(4160)$ and Y(4260) have known widths of $\sim100$~MeV, 
%we can not exclude the possibility that our observed $Z_c$ peaks receive contributions from the excitation of the low energy tail of the  Y(4260) resonance.
%the possibility that our observation of $Z_c(3900)$ at $\psi(4160)$ can arise from the tail of Y(4260) can not be discounted.

This investigation was done using CLEO-c data, and as members of the former CLEO Collaboration we thank it for this privilege. 
This research was supported by the U.S. Department of Energy.


\bibliographystyle{model1a-num-names}

\begin{thebibliography}{00}



\bibitem{besz}  M. Ablikim \textit{et al.} [BES III Collaboration],  Phys. Rev. Lett. \textbf{110}, 252001 (2013).

\bibitem{bellez} Z. Q. Liu \textit{et al.} [Belle Collaboration],  Phys. Rev. Lett. \textbf{110}, 252002 (2013).

%\bibitem{besz}  M. Ablikim \textit{et al.} [BES III Collaboration], \texttt{arXiv:1303.5949[hep-ex]}, published as Phys. Rev. Lett. \textbf{110}, 252001 (2013) while the present Letter was in review.

%\bibitem{bellez} Z. Q. Liu \textit{et al.} [Belle Collaboration], \texttt{arXiv:1304.0121[hep-ex]}, published as Phys. Rev. Lett. \textbf{110}, 252002 (2013) while the present Letter was in review.

\bibitem{belle} Belle Collaboration: R. Mizuk \textit{et al.}, Phys. Rev. D \textbf{78}, 072004 (2008); 
R. Mizuk \textit{et al.}, Phys. Rev. D \textbf{80}, 031104(R) (2009);
A. Bondar \textit{et al.}, Phys. Rev. Lett. \textbf{108}, 122001 (2012).

\bibitem{nuz} A preliminary report of the present results was presented in T.~Xiao \textit{et al.}, \texttt{arXiv:1304.3036[hep-ex]}.



%\bibitem{belle1} R. Mizuk \textit{et al.} [Belle Collaboration], Phys. Rev. D \textbf{78}, 072004 (2008).

%\bibitem{belle2} R. Mizuk \textit{et al.} [Belle Collaboration], Phys. Rev. D \textbf{80}, 031104(R) (2009). 

%\bibitem{belle3} A. Bondar \textit{et al.} [Belle Collaboration], Phys. Rev. Lett. \textbf{108}, 122001 (2012).


\bibitem{cleodetector}
See, for example, S. Dobbs \textit{et al.} [CLEO Collaboration], Phys. Rev. D \textbf{76}, 112001 (2007).


%Y. Kubota \textit{et al.} [CLEO Collaboration], Nucl. Instrum. Meth. A
%\textbf{320}, 66 (1992); M. Artuso \textit{et al.}, Nucl. Instrum.
%Meth. A \textbf{554}, 147 (2005); D. Peterson \textit{et al.}, Nucl.
%Instrum. Meth. A \textbf{478}, 142 (2002).

%\bibitem{tingx}  T.~Xiao \textit{et al.}, Phys. Rev. D \textbf{87}, 057501 (2013).


\bibitem{bonneaumartin} G. Bonneau and F. Martin, Nucl. Phys. B \textbf{27}, 381 (1971).


%\bibitem{lum} S. Dobbs \textit{et al.} [CLEO Collaboration], Phys. Rev. D \textbf{76}, 112001 (2007).

\bibitem{pdg} J. Beringer \textit{et al.} [Particle Data Group], Phys. Rev. D \textbf{86}, 010001 (2012).

%\bibitem{maiani} L. Maiani \textit{et al.}, \texttt{arXiv:1303.6857}.

%%\bibitem{bellez} Z. Q. Liu \textit{et al.} [Belle Collaboration], \texttt{arXiv:1304.0121}.


\end{thebibliography}



\end{document}